\documentclass[12pt, iopfts]{iopart}

 \expandafter\let\csname equation*\endcsname\relax
  \expandafter\let\csname endequation*\endcsname\relax

\usepackage{hyperref}
\usepackage{amsmath}
\usepackage{amsfonts}
\usepackage{amssymb}
\usepackage{xcolor}
\usepackage{graphicx}

  \expandafter\let\csname equation*\endcsname\relax
  \expandafter\let\csname endequation*\endcsname\relax

\newcommand{\ket}[1]{\ensuremath{\left|#1\right\rangle}}
\newcommand{\bracket}[2]{\ensuremath{\left\langle #1 \middle| #2 \right\rangle}}

\newcommand{\eq}[1]{Eq.~(\ref{#1})}
\newcommand{\se}[1]{Sec.~\ref{#1}}

\begin{document}

%% TITLE 
\title[AQO for Associative Memory Recall]{Adiabatic Quantum Optimization for Associative Memory Recall}
\author{H. Seddiqi and T. S. Humble\footnote{MS 6015, Oak Ridge National Laboratory, One Bethel Valley Road, Oak Ridge, TN, 37831-6015 USA}}
\date{today}
\address{Quantum Computing Institute, Oak Ridge National Laboratory, Oak Ridge TN, USA}
\ead{humblets@ornl.gov}
%%%%%%%%%%
%% ABSTRACT
\begin{abstract}
Hopfield networks are a variant of associative memory that recall patterns stored in the couplings of an Ising model. Stored memories are conventionally accessed as fixed points in the network dynamics that correspond to energetic minima of the spin state. We show that memories stored in a Hopfield network may also be recalled by energy minimization using adiabatic quantum optimization (AQO). Numerical simulations of the underlying quantum dynamics allow us to quantify AQO recall accuracy with respect to the number of stored memories and noise in the input key. We investigate AQO performance with respect to how memories are stored in the Ising model according to different learning rules. Our results demonstrate that AQO recall accuracy varies strongly with learning rule, a behavior that is attributed to differences in energy landscapes. Consequently, learning rules offer a family of methods for programming adiabatic quantum optimization that we expect to be useful for characterizing AQO performance.
\end{abstract}

%%%%%%%%%%
%% TITLE
% \maketitle

%%%%%%%%%%%%%%%%%%%%%%%%%%%%%%%%%%%%%%%%%%%%%%%%%%%
% INTRO
%%%%%%%%%%%%%%%%%%%%%%%%%%%%%%%%%%%%%%%%%%%%%%%%%%%
\section{Introduction}
\label{sec:intro}
Content-addressable memory (CAM) is a form of associative memory that recalls information by value \cite{Hopfield1982}. Given an exact or approximate input value, a CAM returns the closest matching key stored in memory. This is in contrast to random access memory (RAM), which returns the value stored at a provided key or address. CAMs are of particular interest for applications tasked to quickly search large databases including, for example, network switching, pattern matching, and machine vision \cite{Pagiamtzis2006}. An auto-associative CAM is a memory in which the key and value are the same and partial knowledge of the input value triggers complete recall of the key.  
\par
Auto-associative CAMs have proven of interest for modeling neural behavior and cognition \cite{Rojas1996}. This is due partly to their properties of operating in massively parallel mode and being robust to noisy input. These features motivated Hopfield to propose a model for an auto-associative CAM based on a network of computational neurons \cite{Hopfield1982, Hopfield1986}. The Hopfield neural network stores memories in the synaptic weights describing the connectivity between the neurons. An initial state of the neural network propagates discretely by updating each neuron based on the synapses and states of the other neurons. Hopfield showed that  memories stored in the network become fixed point attractors under these Markov dynamics. The Hopfield network functions as an auto-associative CAM in which the initial network state represents the input value and the final state represents the recovered key or memory. The memory capacity for a Hopfield network depends strongly on how the synaptic weights are set \cite{Amit1985, Personnaz1986,Storkey1999}. 
\par
The theoretical underpinning of the Hopfield network is a classical Ising model in which each binary neuron is mapped into a spin-1/2 system \cite{Rojas1996}. The synaptic weights define the couplings between these spins and the susceptibility for a neuron to become activated is set by the applied bias. The energy of the Ising model represents a Lyapunov function and stochastic dynamics guarantees convergence to a fixed point attractor in the asymptotic limit \cite{Hopfield1982}. Conventionally, Hopfield networks are formulated in terms of an update rule governed by the Ising energy. However, finding stable points of this Lyapunov function can also be viewed as minimization of the network energy \cite{Schneider2007}. In the case of the Hopfield network, spin configurations that minimize the network energy are fixed point attractors representing stored memories. 
\par
A fundamental concern for accurate memory recall is the likelihood for the network dynamics to converge to the correct memory state. Although stored memories are guaranteed to reside at minima in the network energy, the number of stored memories greatly influences the radius of attraction for each stable point \cite{McEliece1987}. The radius of attraction determines how close (measured by Hamming distance) an initial network state must be in order to converge to a fixed point. As the number of stored memories increases, the radius of attraction for each fixed point decreases due to interference between memories  \cite{Schneider2007}. The initial network state must then start closer to the sought after memory in order to accurately recall it.  Conventional Hopfield networks rely on gradient descent to recover these stable fixed points. However, this method lacks any mechanism for escaping from the local minima that represent interfering memories \cite{Rojas1996}.
\par
In this work, we investigate the recall accuracy of an auto-associative CAM using methods of energy minimization by adiabatic quantum optimization (AQO). AQO represents a novel approach to optimization that leverages quantum computational primitives for minimizing the energy of a system of coupled spin states \cite{Farhi2001,Santoro2002}. In particular, AQO recovers the spin state that corresponds with the global minimum in energy. We formulate memory recall in terms of global energy minimization by AQO in order to avoid the local minima that undermine gradient descent in conventional Hopfield networks. We apply the promise that AQO returns the  global network minimum by investigating how accurately a sought-after-memory can be recalled. As part of the broader adiabatic quantum computing model, AQO has also been investigated for a number of applications, including classification \cite{Neven2008a,Neven2008b}, machine learning \cite{Pudenz2013}, graph theory \cite{Gaitan2012,Hen2012,Bian2013,Gaitan2014}, and protein folding \cite{PerdomoOrtiz2008,PerdomoOrtiz2012} among others \cite{Smelyanskiy2012,Lucas2014,Vinci2014,PerdomoOrtiz2014a,OGorman2014}. In each of these representative applications, the respective problems require reduction first to a discrete optimization problem that is only subsequently mapped into the AQO paradigm. By comparison, we show that memory recall within a Hopfield network is a direct application of AQO. Moreover, this task may be implemented using an Ising model in a transverse field with no reduction in the original problem required \cite{Johnson2011}.
\par
Our analysis is also directed at quantifying the influence that learning rules have on AQO recall accuracy. Although learning rules are well understood to influence memory capacity of Hopfield networks, these rules have not been applied to the study of AQO dynamics. Learning rules define the synaptic couplings that store memories and thus shape the energy landscape of the Ising model. It is an outstanding question to understand how the shape of the energy landscape determines the computational complexity of AQO, and we use these learning rules as a means of comparing performance between different AQO programs that implement the same recall task. This is possible due to the one-to-one correspondence between the Hopfield network and the Ising model. We ensure that the AQO dynamics are always adiabatic by using sufficiently long annealing times in our simulated networks. This enables us to focus on quantifying the relative recall accuracy of AQO under three different learning rules as opposed to questions about adiabaticity. We analyze changes in AQO recall accuracy with respect to the number of stored patterns and type of learning rule employed. We defer to future studies the question of how AQO performs relative to the absolute scaling of the minimum spectral gap. This question for Ising models in a traverse field is presently addressed by many others \cite{Boixo2012, Karimi2012, Ronnow2014, Katzgraber2014}. Our interest is in assessing how learning rules influence recall accuracy in the limit of sufficiently long annealing times. By guaranteeing the adiabatic condition, we avoid trapping in local minima but not interference between memories and the formation of spurious states. We use numerical simulations to quantify the conditions under which AQO may be useful for memory recall.
\par
The use of AQO for performing memory recall in a Hopfield network has been investigated previously by Neigovzen et al.~in the context of pattern recognition \cite{Neigovzen2009}. Specifically, they employed AQO to minimize the energy of a Hopfield network expressed as an Ising Hamiltonian. Neigovzen et al.~performed an experimental demonstration of these ideas using a 2-neuron example in the context of NMR spin-based encoding. Their results confirmed that AQO provided accurate recall for that small network and invited questions as to how  details of the Hopfield network influence performance. Our investigation addresses those questions by quantifying how different network parameters, including size, memories, and learning rules, influence recall accuracy. 
\par
Hopfield networks are tasked with finding an unknown value within an unsorted database, i.e., the network memory. There is a strong connection between this type of tagged search and Grover's search algorithm, which is formulated in terms of a quantum oracle. Previous work by Farhi et al.~ as well as Roland and Cerf using AQO to perform search tasks makes this point clear \cite{Farhi2000, Roland2002}. Both have shown that Grover's search algorithm can be cast in terms of AQO by mapping the oracle operator to the terminal Hamiltonian. A Hopfield neural network using AQO for memory recall is equivalent to these implementations of Grover's search when the oracle expresses a one-memory network. However, a Hopfield network extends the search task to a more general context in which the oracle must discriminate between both tagged and untagged keys. This requires a more complex implementation of the oracle that we find plays a role in overall recall performance. This increase in oracle complexity likely undermine the optimal scaling reported by Roland and Cerf, i.e., $\mathcal{O}(2^{n/2})$, which stores only a single pattern in an $n$-qubit network. Our statistical analyses of multi-memory instances suggest that the optimal annealing schedule is dependent on both the learning rule and the number of stored memories.
\par
In \se{sec:hop}, we define the task of memory recall using a conventional Hopfield network and describe the Hebb, Storkey, and projection learning rules for preparing the synaptic weights. In \se{sec:aqo}, we introduce  adiabatic quantum optimization, its use for memory recall, and the basis for our numerical simulation studies. In \se{sec:recall}, we present results for example instances of Hopfield networks that demonstrate the behavior of AQO for memory recall while in \se{sec:stats} we present calculations of the average recall success for an ensemble of different networks. We present final conclusions in \se{sec:con}.
%%%%%%%%%%%%%%%%%%%%%
%%%%%%%%%%%%%%%%%%%%%
%%% HOPFIELD NETWORKS %%%
%%%%%%%%%%%%%%%%%%%%%
%%%%%%%%%%%%%%%%%%%%%
\section{Hopfield Networks}
\label{sec:hop}
We define a classical Hopfield network of $n$ neurons with each neuron described by a bipolar spin state $z_{j} \in \{\pm 1\}$. Neurons $i$ and $j$ are symmetrically coupled by synaptic weights $w_{ij} = w_{ji}$ while self-connections are not permitted, i.e., $w_{ii} = 0$, to ensure dynamic stability. Different choices for the weights are described below, but in all cases the energy of the network in a spin state $z = (z_{1}, z_{2}, \ldots, z_{n})^{T}$ is
\begin{equation}
\label{eq:energy}
E(z; \theta) = -\frac{1}{2} \sum_{i,j=1}^{n}{z_{i} w_{ij} z_{j}} - \sum_{i =1}^{n}{\theta_i z_{i}},
\end{equation}
with $\theta = (\theta_1, \theta_2, \ldots, \theta_n)^{T}$ and $\theta_i$ the real-valued activation threshold for the $i$-th neuron. This form for the energy represents a classical Ising model in which the spin configuration describes the orientation of the $n$-dimensional system. The dynamics of the Hopfield network are conventionally modeled by the discrete Markov process
\begin{equation}
\label{eq:activation}
z_i = %
    \begin{cases}
        ~~1 & \text{if} ~~\sum_j w_{ij}z_j > \theta_{i} \\
        -1 & \text{otherwise}
    \end{cases}
\end{equation}
where the state of the $i$-th neuron may be updated either in series (asynchronously) or in parallel (synchronously) with all other neurons in the network. The network is initialized in the input state $z_i = z_{0,i}$ and subsequently updated under repeated application of \eq{eq:activation} until it reaches a steady state 
\begin{equation}
\label{eq:update}
z_{i} = \textrm{sign}\bigg(\sum_{j}w_{ij} z_{j}\bigg)
\end{equation}
Steady states of the Hopfield network represent fixed point attractors and are local minima in the energy landscape of \eq{eq:energy} \cite{Rojas1996}. The stable fixed points are set by the choice of the synaptic couplings $w_{ij}$ and the network converges to the memory state closest to the initial state $z_0$. However, the network has a finite capacity to store memories and it is well known that the dynamics converge to a spurious mixture of memories when too many memories are stored \cite{Amit1985, Personnaz1986,Storkey1999}. The emergence of spurious states places a limit on the storage capacity of the Hopfield network that depends on both the interference or overlap between the memories and the learning rule used to set the synaptic weights.
%%%%%%%%%%%%%%%%%%%%%%
%%%%%%%%%%%%%%%%%%%%%%
%% LEARNING RULES %%%%%%%%
%%%%%%%%%%%%%%%%%%%%%%
%%%%%%%%%%%%%%%%%%%%%%
\subsection{Synaptic Learning Rules}
\label{sec:rules}
Learning rules specify how memories are stored in the synaptic weights of a Hopfield network and they play an important role in determining the memory capacity.  The capacity $c_n = p/n$  is the maximum number of patterns $p$ that can be stored in a network of $n$ neurons and then accurately recalled \cite{McEliece1987}.  Different learning rules yield different capacities and we will be interested in understanding how these differences influence performance of the AQO algorithm. Setting the synaptic weights $w_{ij}$ for a Hopfield network is done using a specific choice of learning rule that in turn generates a different Ising model. Learning rules represent a form of unsupervised learning in which the memories are stored in the network without any corrective back-action. We make use of three learning rules that have been found previously to yield different capacities for Hopfield networks in the classical setting.
%%%%%%%%%%%%%%%%%
\subsubsection{Hebb Rule}
\label{sec:hebb}
The Hebb learning rule defines the synaptic weights 
\begin{equation}
\label{eq:hebbrule}
w_{ij} = \frac{1}{n}\sum_{\mu=1}^{p}{\xi_{i}^{\mu}\xi_{j}^{\mu}}
\end{equation}
for a set of $p$ memories $\{ \xi^{1}, \xi^{2}, \ldots, \xi^{p} \}$, each of length $n$ with bipolar elements $\xi_{i}^{\mu}\in\{\pm1\}$. Geometrically, each summand corresponds to the projection of the neuron configuration into the $\mu$-th memory subspace. These projections are orthogonal if all $p$ patterns are mutually orthogonal. More generally, the Hebb rule maps non-orthogonal memory states into overlapping projections. This leads to interference during  memory recall as two or more  correlated memories may both be close to the input state. In the asymptotic limit for the number of neurons, the capacity of the Hebb rule is $c_n = n/2\ln{n}$ under conditions of perfect recall, i.e., no errors in the retrieved state. By comparison, under conditions of imperfect recall the asymptotic capacity is $c_n \approx 0.14$ \cite{Amit1985}. It is worth noting that the Hebb rule is incremental as it is a sum over individual patterns. The rule is also local since the synaptic weights depend only on the value of the adjacent neurons.
%%%%%%%%%%%%%%%%%
\subsubsection{Storkey Rule} 
The Storkey learning rule defines the synaptic weights in an iterative fashion as 
\begin{equation}
\label{eq:stork-rule}
w_{ij}^{\nu} = w_{ij}^{\nu - 1} + \frac{1}{n} \xi_i^{\nu} \xi_j^{\nu} - \frac{1}{n} \xi_i^{\nu} h_{ji}^{\nu} - \frac{1}{n} h_{ij}^{\nu} \xi_{j}^{\nu} 
\end{equation}
where $\xi^{\nu}$ is the memory to be learned in the $\nu$-th iteration for $\nu = 1$ to $p$ and
\begin{equation}
h_{ij}^{\nu} = \sum\limits_{k=1, k\ne i,j} w^{\nu-1}_{ik} \xi_k^{\nu}
\end{equation}
is the local field at the $i$-th neuron \cite{Storkey1999}. The final synaptic weight  storing $p$ memories is given by $w_{ij} = w^{p}_{ij}$.  The Storkey rule is found to  more evenly distribute the fixed points and increases the capacity of the network. The asymptotic Storkey capacity under prefect recall is $n/\sqrt{2\ln{n}}$, which represents an improvement over the Hebb rule. As with the Hebb rule, the Storkey rule is incremental and permits the addition of new memories.
%%%%%%%%%%%%%%%%%
\subsubsection{Projection Rule}
The projection rule defines the synaptic weights for $p$ memories as
\begin{equation}
\label{eq:proj-rule}
w_{ij} = \frac{1}{n} \sum_{\mu,\mu'=1}^{p} \xi^\mu_i C^{-1}_{\mu \mu'} \xi_j^{\mu'}
\end{equation}
where $C_{\mu\mu'} = \frac{1}{n} \sum_{k=1}^n \xi_k^\mu \xi_k^{\mu'}$ is the covariance matrix and $C^{-1}$ is the inverse of $C$. This rule has a theoretical capacity of $n$ for linearly independent patterns and approximately $n/2$ for interfering memories \cite{Personnaz1986,Kanter1987}. The projection rule is neither local nor incremental as adding memories to the network requires resetting each element using knowledge of all other memories. In the limit of orthogonal memories, all three learning rules reduce to the Hebb rule.
%%%%%%%%%%%%%%%%%%%%%%%
%%% RECALL BY AQO %%%%%%%%%
%%%%%%%%%%%%%%%%%%%%%%%
%%%%%%%%%%%%%%%%%%%%%%%
\section{Memory Recall by Adiabatic Quantum Optimization}
\label{sec:aqo}
The learning rules defined in \se{sec:rules} offer different methods for preparing the synaptic weights and the fixed points of a Hopfield network. Conventionally, the network finds those states that satisfy the equilibrium condition of \eq{eq:update} by evolving under the discrete Markov process of \eq{eq:activation}. However, the fixed points of a Hopfield network are also minima of the energy function known as stable fixed points. The stability of these solutions is due to the quadratic form of the energy function $E(z;\theta)$, which is a Lyapunov function that monotonically decreases under updates of network state \cite{Rojas1996}.  As an example, consider that the $k$-th spin in the state $z$ updates, i.e., $z_{k} \rightarrow z_{k}^\prime$. The relative change in  unbiased energy is then
\begin{equation}
\Delta E(z^\prime,z) = - 2 (z_k^\prime - z_k) \sum_{j}{ w_{jk} z_j} \leq 0
\end{equation}
The sign of the summation always correlates with the change in the spin state, cf. \eq{eq:activation}. Thus, network energy never increases with respect to updates in the state $z$. More importantly, the Lyapunov stability of the Hopfield network guarantees that the stochastic dynamics converge to fixed points representing stored memories.
\par
As an alternative to fixed point convergence under stochastic update, we apply the principle of optimization for finding the global minima of the energy function and for recalling a stored memory. We are motivated by the stability analysis of the Ising model under Markov dynamics, which guarantees that memories represent fixed point attractors and, more importantly, energy minima. Our formulation uses the same synaptic weight matrix and underlying Ising model of a conventional Hopfield network. However, we set the activation thresholds $\theta_i$ in place of initializing the network to a known initial state $z_0$. This feature casts recovery of an unknown memory in terms of minimizing the energy of the network. We formally define the energy minimization condition as
\begin{equation}
\label{eq:minz}
z = \arg\min_{z'} E(z'; \theta).
\end{equation}
in which the vector $\theta$ represents the activation thresholds $\theta_i = \Gamma  z_{0,i}$ and $\Gamma$ is an energy scale for the  applied bias. The activation threshold $\theta$ serves as an energetic bias towards network states that best match the input $z_0$. The behavior for a classical Hopfield network is recovered by initializing the state of all neurons to an indeterminate value, i.e., $z_i = 0$, and using the first update to prepare the state $z_0$. 
\par
In the absence of any bias, finding the global minima of $E(z,0)$ is equivalent to computing the lowest energy eigenstates of the synaptic weight matrix $w_{ij}$ with the constraint $z_i\in\{\pm1\}$ (indeterminate values are not valid output states). Due to the symmetry of the unbiased energy, the complement of each memory is also an eigenstate. If the network stores $p$ memories, then the ground state manifold is $2p$ degenerate subspace. However, the presence of a non-zero bias breaks this symmetry and leads to a lower energy for only one memory state relative to the other stored memories. 
\par
In the presence of bias, global minimization of $E(z,\theta)$ returns the spin configuration that encodes a recalled memory. The promise that the encoded memory is a global minimum depends on several factors. First, if the applied bias is too large then the input state itself becomes a fixed point and the global minimum becomes $z_{0}$. This behavior is unwanted since it does not confirm whether the input or its closest match were part of the memory. This effect can be detected by decreasing $\Gamma$ and monitoring changes in the recall. However, we can also compute an upper bound on $\Gamma$ by comparing network energies of a memory state $\xi^{k}$ with a non-memory state $z_0$, e.g., for the projection rule
\begin{equation}
\Gamma < \frac{\sum_{i,j}{\xi_{i}^{k} w_{ij} \xi_{j}^{k} - z_{0,i} w_{ij} z_{0,j}}}{2(n - \sum_{i}{z_{0,i} \xi_{i}^{k}})}
\end{equation}
ensures that the network does not become over biased. In the limit that the memories are orthogonal to each other as well as the input key, this reduces to the result $\Gamma <1/(2n)$ previously noted by Neigovzen et al.~\cite{Neigovzen2009}.
\par
Interference between memories prevents their discrimination when insufficient knowledge about the sought-after memory is provided. The number of memories stored in the network may also exceed the network capacity and lead to erroneous recall results. As an example, perfect recall is observed when using the Hebb rule in a classical network storing $p$ orthogonal memories provided $p \leq n$, since there is no interference in these non-overlapping states. However, the capacity for non-orthogonal memories is much lower and varies with learning rule, as described above. In our optimization paradigm, interference manifests as degeneracy in the ground state manifold. These degeneracies are formed from superpositions of stored memory states and the applied bias. These states are valid energetic minima that correspond to the aforementioned spurious states. Differences between learning rules seek to remove the presence of spurious states while also increasing the network capacity.
%%%%%%%%%%%%%%%%%%%%%%
%%%%%%%%%%%%%%%%%%%%%%
%% AQO ALGORITHM%%%%%%%%%
%%%%%%%%%%%%%%%%%%%%%%
%%%%%%%%%%%%%%%%%%%%%%
\subsection{Adiabatic Quantum Optimization Algorithm}
\label{sec:aqoa}
Adiabatic quantum optimization (AQO) is based on the principle of adiabatically evolving the ground state of an initial well-known Hamiltonian to the unknown ground state of a final Hamiltonian. By defining the final Hamiltonian  in terms of the Ising model representing a Hopfield network, we use  AQO to recover the ground state expressing a stored memory. The Ising model for AQO will use the same synaptic weights and activation thresholds discussed in \se{sec:hop} for the Hopfield network. The recall operation begins by preparing a register of $n$ spin-1/2 quantum systems (qubits) in a superposition of all possible network states and adiabatically evolving the register state towards the final Ising Hamiltonian. Assuming the adiabatic condition has remained satisfied, the qubit register is prepared in the ground state of the Ising Hamiltonian. Upon completion of the evolution, each qubit in the register is then measured and the resulting string of bits is interpreted as the network state, $z_{i}$.
\par
Formally, we consider a time-dependent Hamiltonian 
\begin{equation}
\label{ht}
H(t) = A(t) H_0 + B(t) H_{1}
\end{equation}
with piece-wise continuous annealing schedules $A(t)$ and $B(t)$ that satisfy $A(0) = 1, B(0) = 0$ and $A(T) = 1, B(T) = 1$. Together, the initial Hamiltonian
\begin{equation}
\label{h0}
H_0 = -\sum_{i}^{n}{X_{i}}
\end{equation}
and the final Hamiltonian 
\begin{equation}
\label{h1}
H_1 = -\sum_{i,j}^{n}{J_{ij} Z_{i} Z_{j}}-\sum_{i}^{n}{h_{i} Z_{i}}
\end{equation}
represent an Ising model in a transverse field. In the latter equations, the Pauli $Z_{i}$ and $X_{i}$ operators act on the $i$-th qubit while the constants $h_i$ and $J_{ij}$ denote the qubit bias and coupling, respectively. Of course, the latter quantities are exactly the activation threshold and synaptic weights of the Hopfield network, i.e., $h_i = \theta_i$ and $J_{ij} = w_{ij}$, and we use the symbols interchangeably. We choose the computational basis in terms of tensor product states of the $+1$ and $-1$ eigenstates of operators $Z_i$ denoted as $\ket{0}$ and $\ket{1}$, respectively. In this basis, the correspondence between the binary spin label $s_i \in \{0,1\}$ and the bipolar spin configuration label is $z_i = 2s_i-1$.
\par
The quantum state of an $n$-qubit register is prepared  at time $t=0$ in the ground state of $H_0$, 
\begin{equation}
\label{eq:psi0}
\ket{\psi(t=0)} =\frac{1}{\sqrt{2^n}} \sum_{x=0}^{2^n-1}{\ket{s}},
\end{equation}
with $\ket{s} = \ket{s_1}\otimes\ket{s_2}\ldots\otimes\ket{s_{n}}$ and
\begin{equation}
s = \sum_{i=1}^{n}{s_i 2^{i-1}},\hspace{1cm}s_i\in\{0,1\}
\end{equation}
the binary expansion of the state label $s$. The register state $\psi(t)$ evolves under the Schrodinger equation
\begin{equation}
\label{tdse}
i  \frac{d \ket{\psi(t)}}{dt} = H(t) \ket{\psi(t)}
\end{equation}
from the initial time $0$ to a final time $T$. We  set $\hbar = 1$. The time scale $T$ is chosen so that changes in the register state $\psi(t)$ are slow (adiabatic) relative to the inverse of the minimum energy gap of $H(t)$, which has instantaneous eigenspectrum
\begin{equation}
H(t)\ket{\varphi_{i}(t)} = E_{i}(t) \ket{\varphi_{i}(t)}\hspace{1cm}i=1\textrm{ to } 2^{n},
\end{equation}
where the $i$-th eigenstate $\varphi_i$ has energy $E_i$. The minimum energy gap $\Delta_{min}$ is defined as the smallest energy difference between the instantaneous ground state manifold and those excited states that do not terminate as a ground state. Provided the time scale $T\gg\Delta_{min}^{-\alpha}$ for  $\alpha=2,3$, then the register typically remains in the ground state of the instantaneous Hamiltonian and evolution to the time $T$ prepares the ground state of $H(T) = H_1$. However, the exact scaling for the minimal $T$ with respect to Ising model size and parametrization is an open question.
\par
After preparation of the final register state $\psi(T)$, each qubit is measured in the computational basis. Because the final Hamiltonian $H_1$ is diagonal in the computational basis, measurement results represent the prepared (ground) state. The measurements may be directly related to a valid spin configuration of the Hopfield network. The state of the $i$-th qubit is measured in the $Z_i$ basis and the resulting label $z_i$ is the corresponding spin configuration for the $i$-th neuron.
\par
Regarding execution time, the average-case time complexity for the AQO algorithm is currently observed to require $T \gg \Delta_{min}^{-\alpha}$ with $\alpha=2,3$ in order to recover the global minimum with negligible error. The scaling of energy gap $\Delta_{min}$ with respect to $n$, however, is currently poorly understood except in a few cases. For example, worst-case behavior predicts a gap that shrinks exponentially with increasing $n$, whereas polynomial scaling is observed in some narrow instances. By comparison, the algorithmic complexity for stochastic update in \eq{eq:update} is dominated by the matrix-vector multiply. Assuming the classical operations are directly proportional to time, the execution time for stochastic update scales as $O(n^2)$. If $\Delta_{min}$ scaled as $1/n$, then AQO would at best have the same time scaling as stochastic update. Such weak scaling of the energy gap is unlikely for the average case, and it is far more likely that AQO provides a slowdown relative to gradient descent. This is because AQO makes a stronger promise than the stochastic update rule in \eq{eq:update}, i.e., the latter only finds a local stable fixed point.
%%%%%%%%%%%%%%%%%%%%%%%
%%%%%%%%%%%%%%%%%%%%%%%
%% MEMORY RECALL ACCURACY %%
%%%%%%%%%%%%%%%%%%%%%%%
%%%%%%%%%%%%%%%%%%%%%%%
\subsection{AQO Recall Accuracy}
The accuracy with which a memory is recalled using the AQO algorithm can be measured in terms of the probability that the correct (expected) network state is recovered. We define a measure of the probabilistic recall success as
\begin{equation}
\label{eq:fx}
f_{x} =\left\{
\begin{array}{c}
1, \> P_{ans} \geq x \vspace{1em} \\
0, \> P_{ans} < x
\end{array}
\right.
\end{equation}
where $P_{ans}$ is the probability to recover the correct memory and $x \in [0,1]$ is the threshold probability. Denoting the correct memory state as $\phi_{ans}$, the probability to recover the correct memory can be computed from the simulated register state as
\begin{equation}
P_{ans} = |\bracket{\phi_{ans}}{\psi(T)}|^2
\end{equation}
We  assume in this analysis that the register state is a pure state and therefore neglect sources of noise including finite temperature and external couplings.
\par
From this definition for probabilistic success, we consider average success for an ensemble of $N$ problem instances as
\begin{equation}
\label{eq:fsucc}
\langle f_{x} \rangle = \frac{1}{N}\sum_{i=1}^{N}{f_{x}^{i}},
\end{equation}
where $f_{x}^{i}$ represents the probability for success of the $i$-th problem instance of $n$ neurons storing $p$ memories. This is a binomial distribution with variance $\langle \Delta f_{x} \rangle = \langle f_{x} \rangle (1-\langle f_{x} \rangle)$. We use the statistic $\langle f_{x} \rangle$ to characterize accuracy for the ensemble of simulated recall operations.
\par
We use several tests of recall accuracy to characterize each learning rule. First, we quantify the recall success with respect to the applied bias when recalling a state known to be stored in the network. This removes any uncertainty (noise) in the input $z_0$. Second, we quantify recall as the failure rate when the input $z_0$ is noisy. This tests the ability for the network to discriminate noisy input from unknown memories. We quantify noise in terms of Hamming distance of the input state from the expected memory state. We perform these tests for all three learning rules and variable numbers of stored memories. 
%%%%%%%%%%%%%%%%%%%%%%
%%%%%%%%%%
\subsection{Numerical Simulations of the AQO Algorithm}
\label{sec:sim}
We use numerical simulations of the time-dependent Schrodinger equation in \eq{tdse} to compute the register state $\psi(T)$ prepared by the AQO algorithm. These simulations provide the information needed to calculate the probabilistic success $f_x$ as well as the average success with respect to network size and learning rule. Our methods are restricted to pure-state simulations, which provide an idealized environment for the AQO algorithm and permit our analysis to emphasize how learning rules influence success via changes to the Ising model.
\par
\begin{figure}
\begin{center}
\includegraphics[width=17cm]{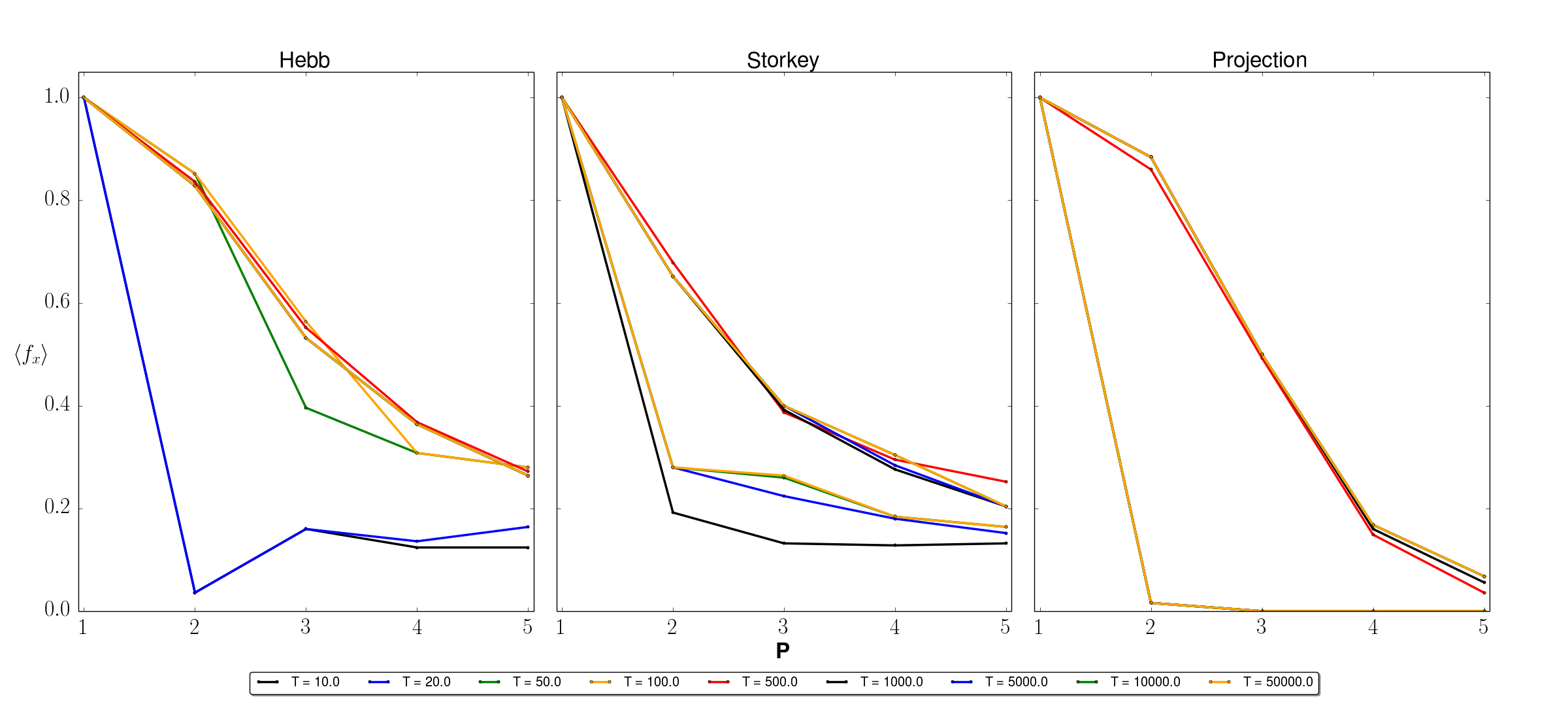}
\end{center}
\textbf{\refstepcounter{figure}
\label{fig:fx_vs_p_multi} Figure \arabic{figure}. }{Average probability to recall the correct memory with respect to number of input patterns for $n=5$ neurons. Each panel plots recall success with respect to the number of stored patterns $p=1,2,3,4$ and $5$ for a different learning rule and applied bias (Hebb, $\Gamma=0.5$; Storkey $\Gamma=0.15$; projection, $\Gamma=0.15$). Each line corresponds to a different annealing time $T=10, 20, 50, 100, 500, 1000,  5000, 10000$ and $50000$. The computed recall success converges as $T$ increases with upper bounds given by (Hebb) $T<50$, (Storkey) $T<500$, (projection) $T<500$. Note that results reported in other figures used an annealing time much longer than these upper bounds, i.e., $T=1000$. }
\end{figure}

Our numerical methods make use of a first-order Magnus expansion of the time-evolution operator
\begin{equation}
\label{teo}
U(t_{j+1},t_j) = \exp\left[-i\int_{t_j}^{t_{j+1}}{ H(\tau) d \tau} \right]
\end{equation}
over the interval $[t_j, t_{j+1}]$ for $j =0\textrm{ to } j_{\textrm{max}}-1$. The use of a first-order approximation is justified by limiting our simulations to annealing times $T$ that produce states well approximated by the ground state. We confirm this approximation by testing the convergence of the ground state population with respect to $T$, cf. Fig.~\ref{fig:fx_vs_p_multi}. Our simulations use a uniform time step $\Delta t = t_{j+1} - t_{j}$ such that $T = j_{\textrm{max}} \Delta t$.  Starting from the initial state \eq{eq:psi0}, an intermediate state is generated from the series of time evolution operators
\begin{equation}
\ket{\psi(t_{j'})} = \prod_{j=0}^{j'-1}{U(t_{j+1}, t_{j})}\ket{\psi(0)}
\end{equation}
In these calculations, the action of the $j$th time-evolution operator onto the appropriate state vector is calculated directly \cite{Al-Mohy2011, Higham2010}. The simulation code is available for download \cite{ADIAQC}. In our simulations, we use annealing schedules $A(t) = 1-t/T$ and $B(t) = t/T$, and we do not place any constraints on the qubit connectivity or the coupling precision. Simulated problem instances are detailed below but in general input parameters include the number of neurons $n$, the number of stored patterns $p$, the applied learning rule (Hebb, Storkey, or projection), the annealing time $T$, the applied bias $\Gamma$ and the input key $z_0$. The large number of parametrized simulations has limited our problem instances to only a few neurons.
%%%%%%%%%%%%%%%
%%%%%%%%%%%%%%%%
%%%%%%%%%%%%%%%%
%%%%%%%%%%%%%%%%
\section{Recall Instances}
\label{sec:recall}
\begin{figure}[h]
\begin{center}
\includegraphics[width=12cm]{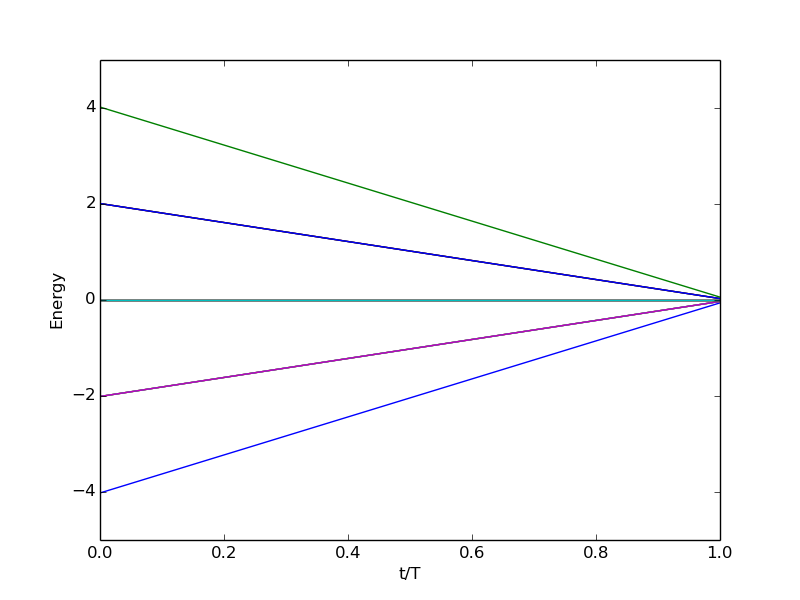}
\end{center}
\textbf{\refstepcounter{figure}
\label{fig:n4ortho} Figure \arabic{figure}. }{Time-dependent eigenspectrum for $p=4$ orthogonal memories stored in a network of $n = 4$ neurons in the absence of bias, $\theta_i = 0$. For orthogonal memories, the spectrum is the same for the Hebb, Storkey, and projection learning rules.}
\end{figure}

\begin{figure}
\begin{center}
\includegraphics[width=12cm]{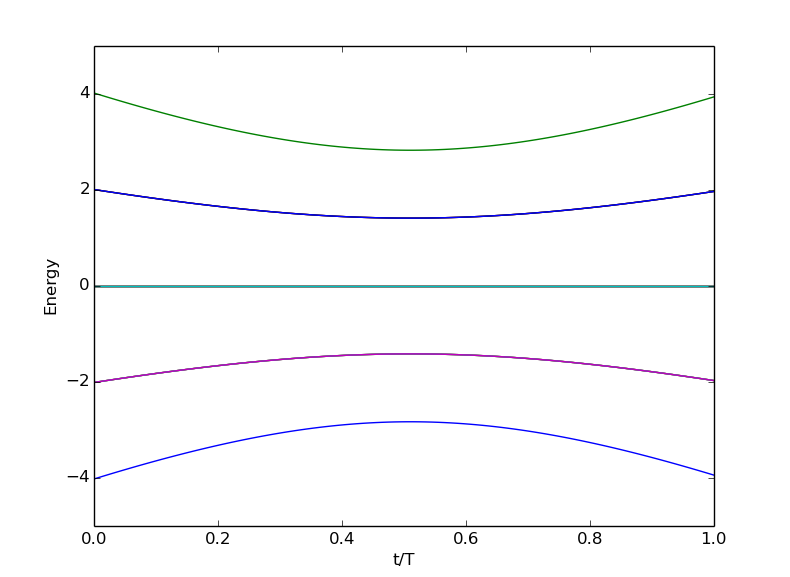}
\end{center}
\textbf{\refstepcounter{figure}
\label{fig:n4p4G1OrthoRecall} Figure \arabic{figure}. }{Time-dependent eigenspectrum for $p=4$ orthogonal memories stored in a network of $n = 4$ neurons in the presence of bias. We define $\theta$ in terms of the first input memory and $\Gamma = 1$. For orthogonal memories, the spectrum is the same for the Hebb, Storkey, and projection learning rules.}
\end{figure}

We first present some example instances to demonstrate AQO behavior during memory recall for different learning rules. We begin by considering the case of $p$ orthogonal memories. A convenient source of orthogonal bipolar states is the $n$-dimensional Hadamard matrix for $n = 2^k$, whose unnormalized columns are orthogonal with respect to the usual inner product. We use these memories to prepare the synaptic weights and corresponding Ising Hamiltonians. Orthogonal memories are a special case for which all learning rules prepare the same synaptic weights. In the absence of any bias ($\theta_i = 0$), we expect recall to recover each of the $p$ encoded memories with uniform probability. The quadratic symmetry of the energy in \eq{eq:energy} also makes the complement of each memory state a valid fixed point. This implies a total ground state degeneracy of $2p$ in the absence of bias. An example of the time-dependent spectral behavior is shown in Fig.~\ref{fig:n4ortho} for the case $p=n=4$, and all the eigenstates converge to a single ground state energy. The same case but with $\theta$ set to the first memory  and $\Gamma = 1$ is shown in Fig.~\ref{fig:n4p4G1OrthoRecall}. The presence of the bias removes the ground state degeneracy and, not apparent from the figure, the prepared ground state matches the biased input state.
\par
We next consider an instance of non-orthogonal memories defined to have an non-zero inner product between pairs of memories. Interference is expected to cause failure during recall when the applied bias is insufficient to distinguish between similar states. With $p=n=4$, we use the memory set 
\par
\begin{equation}
\label{eq:n4randmem}
\Sigma = 
\left(
  \begin{array}{cccc}
    +1 & -1 & -1 & +1 \\
    -1 & +1 & -1 & -1 \\
    +1 & +1 & +1 & +1 \\
    -1 & +1 & +1 & +1
  \end{array}
\right)
\end{equation}
where columns 1, 2, and 3 overlap while columns 2 and 4 are orthogonal. We use an input state $z_{0} =  (1, -1, 1, -1)$ that most  matches the first memory $\Sigma_{i,1}$. For these simulations, we found the annealing time $T= 1000$ was sufficiently long to yield convergence in the prepared quantum state. Both time and energy are expressed in arbitrary units since all calculated quantities are independent of the absolute energy scale of the Ising model.
\par
\begin{figure}
\begin{center}
\includegraphics[width=12cm]{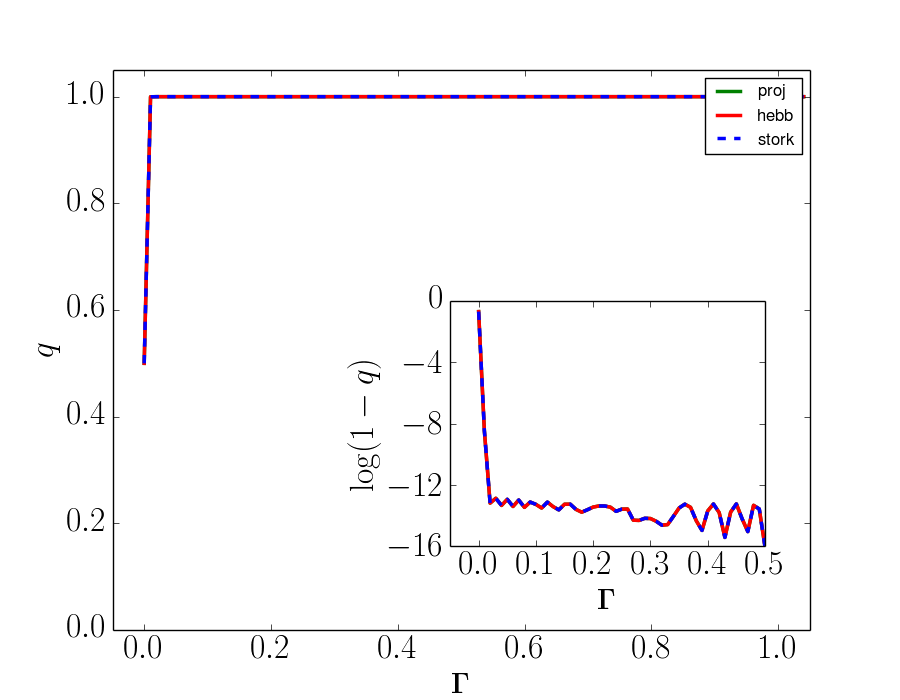}
\end{center}
\textbf{\refstepcounter{figure}
\label{fig:n4randp1} Figure \arabic{figure}. }{Probability $q$ to recall the correct memory with respect to applied bias $\Gamma$ for $n=4$ neurons and $p=1$ memory from the set in \eq{eq:n4randmem}. All three learning rules coincide in behavior and provide unit recall success for any amount of applied bias. Inset: a semi-log plot showing how recall error decreases with applied bias. Numerical noise dominates the inset plot beyond $\Gamma \approx 0.02$}
\end{figure}
\begin{figure}
\begin{center}
\includegraphics[width=12cm]{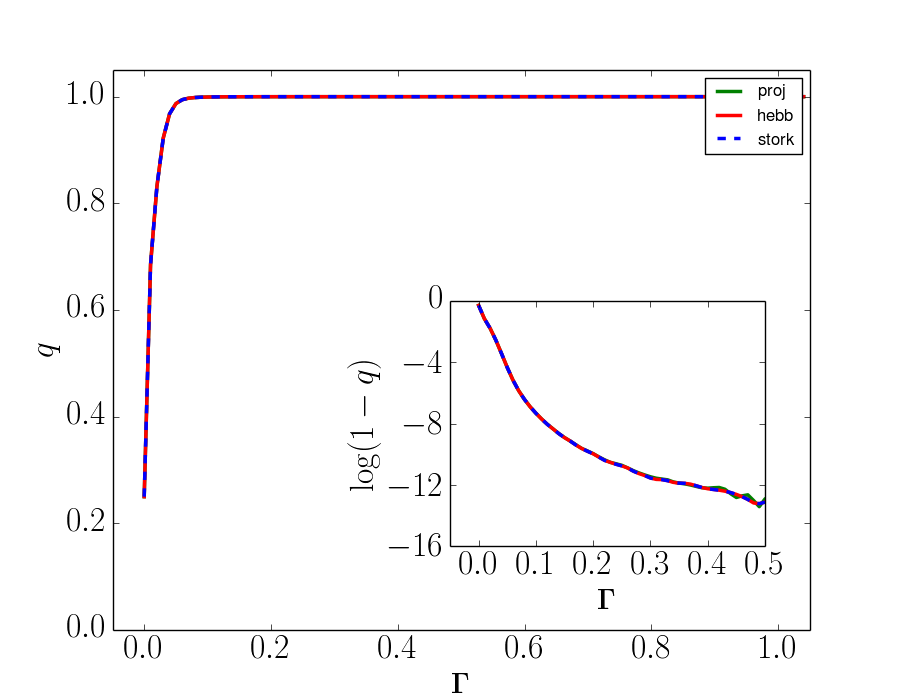}
\end{center}
\textbf{\refstepcounter{figure}
\label{fig:n4randp2} Figure \arabic{figure}. }{Probability $q$ to recall the correct memory with respect to applied bias $\Gamma$ for $n=4$ neurons and $p=2$ memories from the set in \eq{eq:n4randmem}. All three learning rules coincide and show unit recall success for $\Gamma > 0.10$. Inset: a semi-log plot showing how recall error decreases with applied bias. Numerical noise dominates the inset plot beyond $\Gamma \approx 0.4$}
\end{figure}
\begin{figure}
\begin{center}
\includegraphics[width=12cm]{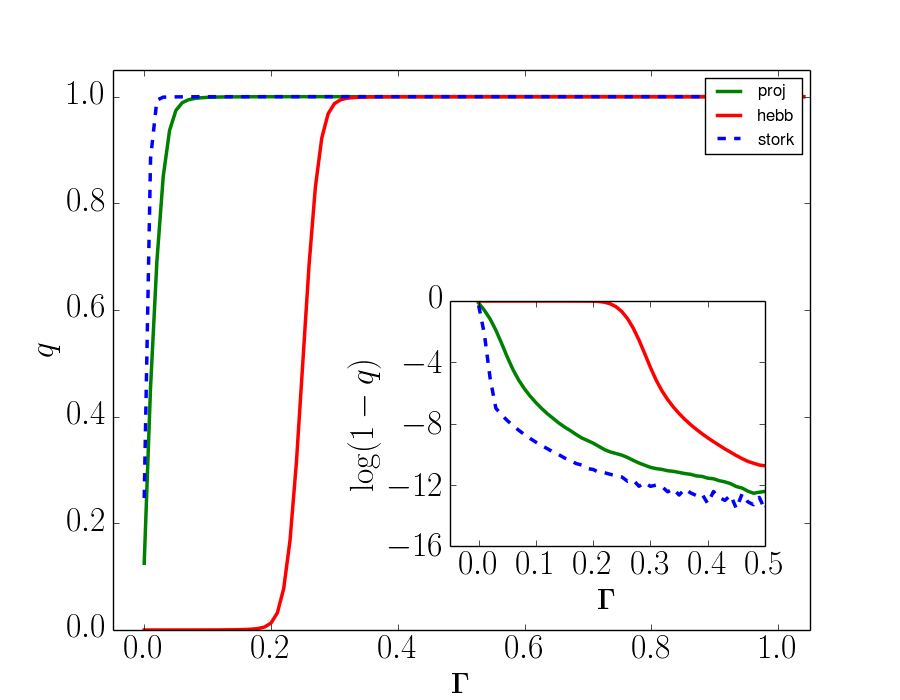}
\end{center}
\textbf{\refstepcounter{figure}
\label{fig:n4randp3} Figure \arabic{figure}. }{Probability $q$ to recall the correct memory with respect to applied bias $\Gamma$ for $n=4$ neurons and $p=3$ memories from the set in \eq{eq:n4randmem}. The Hebb rule has strong dependency on $\Gamma$ due to memory interference while the Storkey and projection rules accommodate interference better. Inset: a semi-log plot showing how recall error decreases with applied bias. Numerical noise dominates the inset plot beyond $\Gamma \approx 0.5$}
\end{figure}
\begin{figure}
\begin{center}
\includegraphics[width=12cm]{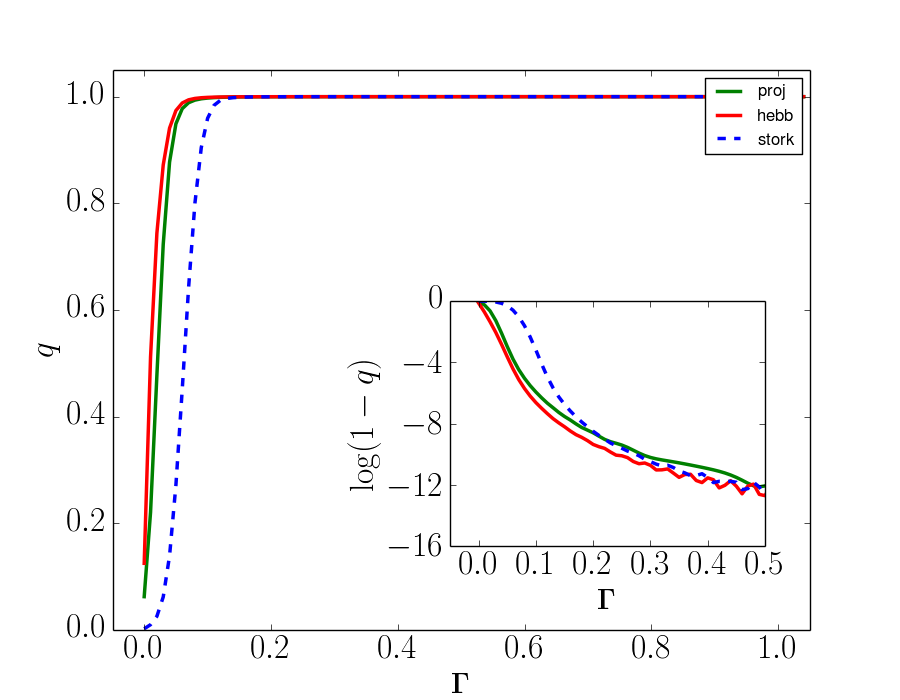}
\end{center}
\textbf{\refstepcounter{figure}
\label{fig:n4randp4} Figure \arabic{figure}. }{Recall probability $q$ with respect to applied bias $\Gamma$ for $n=4$ neurons and $p=4$ memories from the set in \eq{eq:n4randmem}. The Hebb rule is least sensitive to the applied bias, and nearly the same as the projection rule, while the Storkey rule becomes more sensitive to applied bias. Inset: a semi-log plot showing how recall error decreases with applied bias. Numerical noise dominates the inset plot beyond $\Gamma \approx 0.3$ }
\end{figure}

Figures~\ref{fig:n4randp1} through \ref{fig:n4randp4} plot the probability $q=|\langle \varphi_\textrm{ans}|\psi \rangle |^2 $ to successfully recall the answer state $\varphi_\textrm{ans}$ as a function of the applied bias $\Gamma \in [0, 1]$. The inset for each figure shows the semi-log plot of recall error under the same conditions. Recall probability $q$ varies with input bias, number of memories, and learning rule. For $p = 1$, there is only one memory stored in the network and any non-zero bias distinguishes between the memory and its complement. Similarly, all three rules behave the same for the case of $p=2$ in Fig.~\ref{fig:n4randp2}. This is because there are not significant energetic differences between the rules using the first two memories above. The Hebb and Storkey rules coincide exactly, while the projection rule is identical only for the lowest energy eigenstate. However, for the case of $p=3$ in Fig.~\ref{fig:n4randp3}, there is a distinction between all three rules. The answer state probability using the projection rule is nearly the same as was observed for fewer memories while the Hebb rule shows a shift to larger bias. This shift is due to the added memory creating an energy basin that is lower than the unbiased answer state. Larger bias must be applied to lower the answer state below that of the new memory. In contrast, the Storkey shifts to smaller bias as a result of memory addition. This is because the Storkey rule attempts to mitigate interference by using the local field calculation. However, with the addition of another memory, $p=4$, the Hebb rule becomes more evenly distributed in energy across the degenerate memory states while the Storkey rule shows a slight shift to larger bias and the projection rule again remains unchanged. Differences in the recall errors are readily seen from the semi-log plots inset in the figures. The slopes of these lines highlight that each learning rule has a different sensitivity to $\Gamma$. Note that the inset plots show oscillations in the recall error when it is less than about $10^{-12}$; this is due to the finite precision of the numerical simulations. 
%%%%%%%%%%
%%%%%%%%%%
\section{Statistical Recall Behavior}
\label{sec:stats}
Our results for recall success of individual Hopfield networks indicate a large degree of variability in performance with respect to the stored memory states. We have found it useful to average performance across a range of problem instances. Under these circumstances, we use the average success probability defined by \eq{eq:fsucc} to quantify the relative performance of each learning rule in terms of neurons $n$, memories $p$, and bias $\Gamma$. As noted earlier, these statistics correspond to a binomial distribution with parameter $\langle f_x \rangle$.
\par
We first investigate average AQO recall behavior with respect to the bias $\Gamma$. An ensemble of problem instances is constructed for $n = 5$ neurons in which each instance consists of $p$ memories with elements sampled uniform random from $\{\pm1\}$. Among the $p$ memories, one is selected as the answer state while all other memories are distinct from the answer state. The selected answer state is then chosen as the input state. This defines the activation threshold $\theta = \Gamma z_0$ for some choice of $\Gamma$. The simulation computes the full quantum state using an annealing time $T = 1000$. The probability to occupy the expected answer state is then computed using \eq{eq:fx} with a threshold $x = 2/3$. The exact value of $x$ is not expected to be significant provided it is above the probability for a uniform superposition.
\par
\begin{figure}
\begin{center}
\includegraphics[width=17cm]{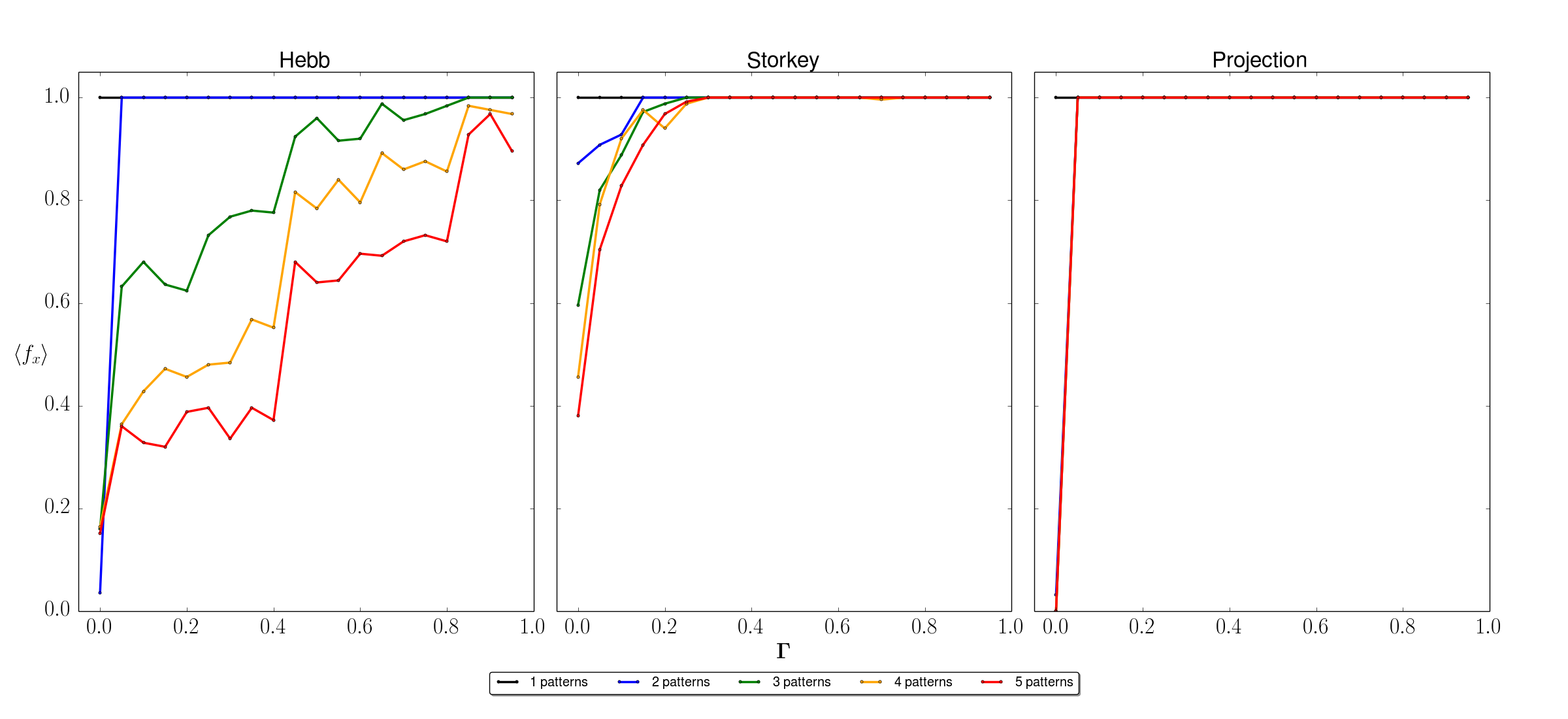}
\end{center}
\textbf{\refstepcounter{figure}
\label{fig:exp3_var_gamma_n5}Figure \arabic{figure}. }{Average probability to recall the correct memory with respect to applied bias. The average is taken over an ensemble of network instances with $n=5$ neurons and $p=1,2,3,4$, or $5$ stored memories. Each panel plots the average recall success of a learning rule (Hebb, Storkey, projection) using an input state that is Hamming distance 0 from a stored memory.}
\end{figure}

Figure~\ref{fig:exp3_var_gamma_n5} shows the average recall success for recovering the answer state as the bias $\Gamma$ increases from 0 to 1. Each panel represents the results of a single learning rule and each line represents a network storing $p=1,2,3,4$ or $5$ memories. We find that each learning rule exhibits a distinct behavior with respect to recall accuracy. For the Hebb rule there is a step-wise decrease in success as the number of memories increases, indicating greater interference during recall. A much weakened version of this dependency is seen for the Storkey rule at values of $\Gamma$ below about 0.25. Above this threshold the Storkey rule recovers unit success for every memory set. The projection rule demonstrates a very different behavior; unit success is seen in every case for any non-zero value of $\Gamma$. Unlike the Hebb rule, there is a complete lack of interference during recall. The plots in Fig.~\ref{fig:exp3_var_gamma_n5} indicate when the prepared ground state has greater than $2/3$ probability to be in the answer state given an input that matches a memory. The better performance of both the Storkey and projection rules is a result of how they exploit correlations between memories. Both rules effectively raise the energy barrier between fixed stable points, while the Hebb rule preserves this interference. As the number of memories increases, so does the interference within the the typical problem instance. This behavior is underscored by the strong dependence of the Hebb rule on the number of stored memories $p$.
\par
\begin{figure}
\begin{center}
\includegraphics[width=17cm]{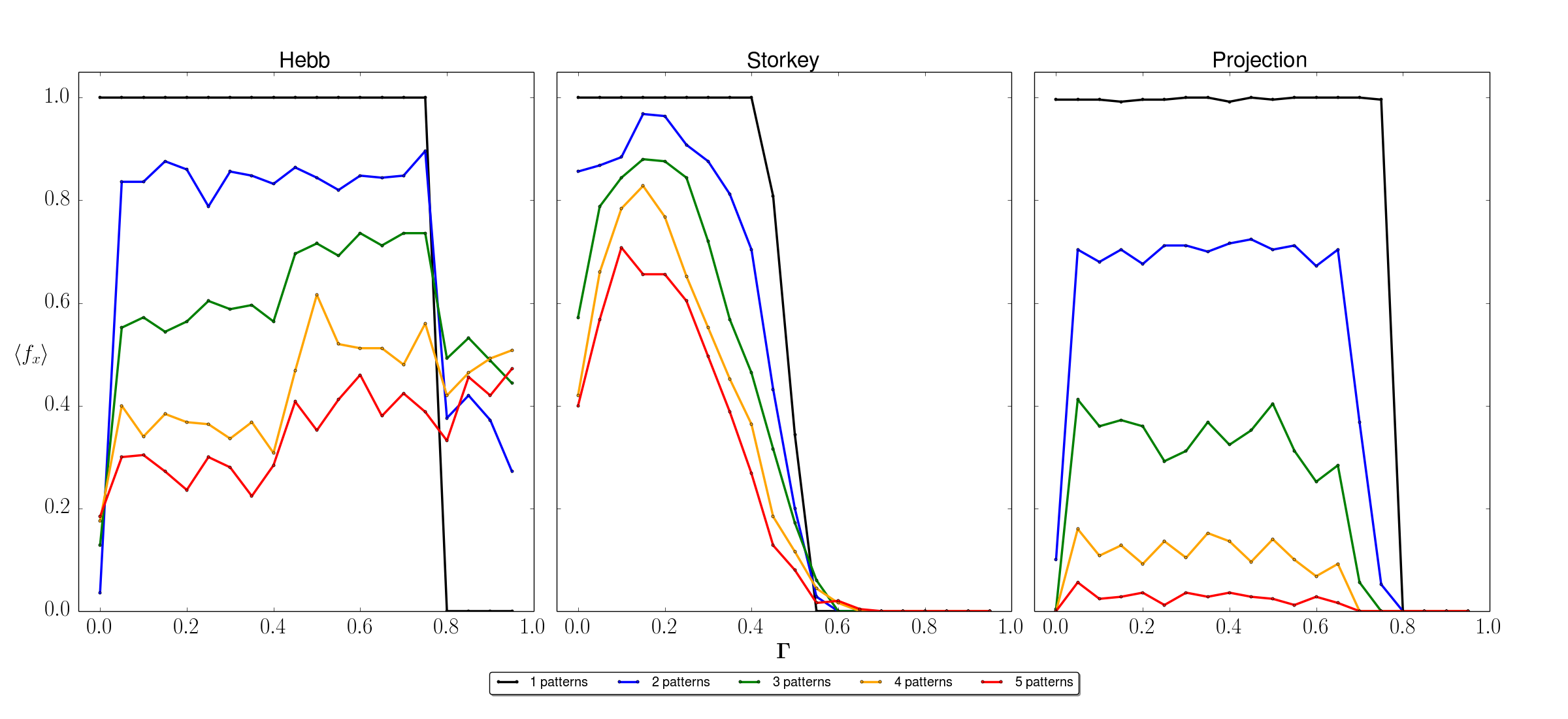}
\end{center}
\textbf{\refstepcounter{figure}
\label{fig:var_gamma_n5} Figure \arabic{figure}. }{Average probability to recall the correct memory with respect to applied bias. The average is taken over an ensemble of network instances with $n=5$ neurons and $p=1,2,3,4$, or $5$ stored memories. Each panel plots the average recall success of a learning rule (Hebb, Storkey, projection) using an input state that is Hamming distance 1 from a stored memory. For sufficiently large $\Gamma$, the recall success drops because the input state becomes the global minimum of the network.}
\end{figure}

We also investigate AQO recall accuracy when a noisy input state is provided. As before, we construct an ensemble of problem instances for $n = 5$ neurons in which each instance consists of $p$ memories with elements sampled uniform random from $\{\pm1\}$. We modify the procedure by selecting one memory as the answer state while creating all other memories at least Hamming distance 2 away from this answer state. We select an input state that is Hamming distance 1 away from the answer state by randomly flipping a bit in the answer state. This  construction of the memory set ensures that the noisy input state is closest to the answer state. In Fig.~\ref{fig:var_gamma_n5}, the recall success for the noisy input test is plotted with respect to $\Gamma$ and number of memories for each learning rule. For the Hebb rule, there is again a step-wise decrease in recall success as the number of stored memories is increased. This behavior indicates that the energy basin representing the answer state is not narrow with respect to the Hamming distance between spin configurations. As the stored memories increase, there is a greater chance that the applied bias lowers the energy of non-answer states. As the bias $\Gamma$ increases beyond about 0.75, the input state is over biased. This leads to a recall accuracy of about $50\%$ independent of stored memories. The Storkey rule exhibits a different behavior with respect to noisy input. Recall accuracy again decreases with the addition of new memories but much more weakly than was observed with the Hebb rule. The recall success also tends to vanish as the bias is increased. These differences underlie the fact that the Storkey rule distributes  stored memories better than Hebb, such that an over-biased input is well separated from the expected answer state. Recall accuracy with the projection rule also vanishes for sufficiently strong bias due to the well-separated memory states. However, there is a much stronger dependence on recall accuracy with respect to the number of stored memories.
\par
\begin{figure}
\begin{center}
\includegraphics[width=17cm]{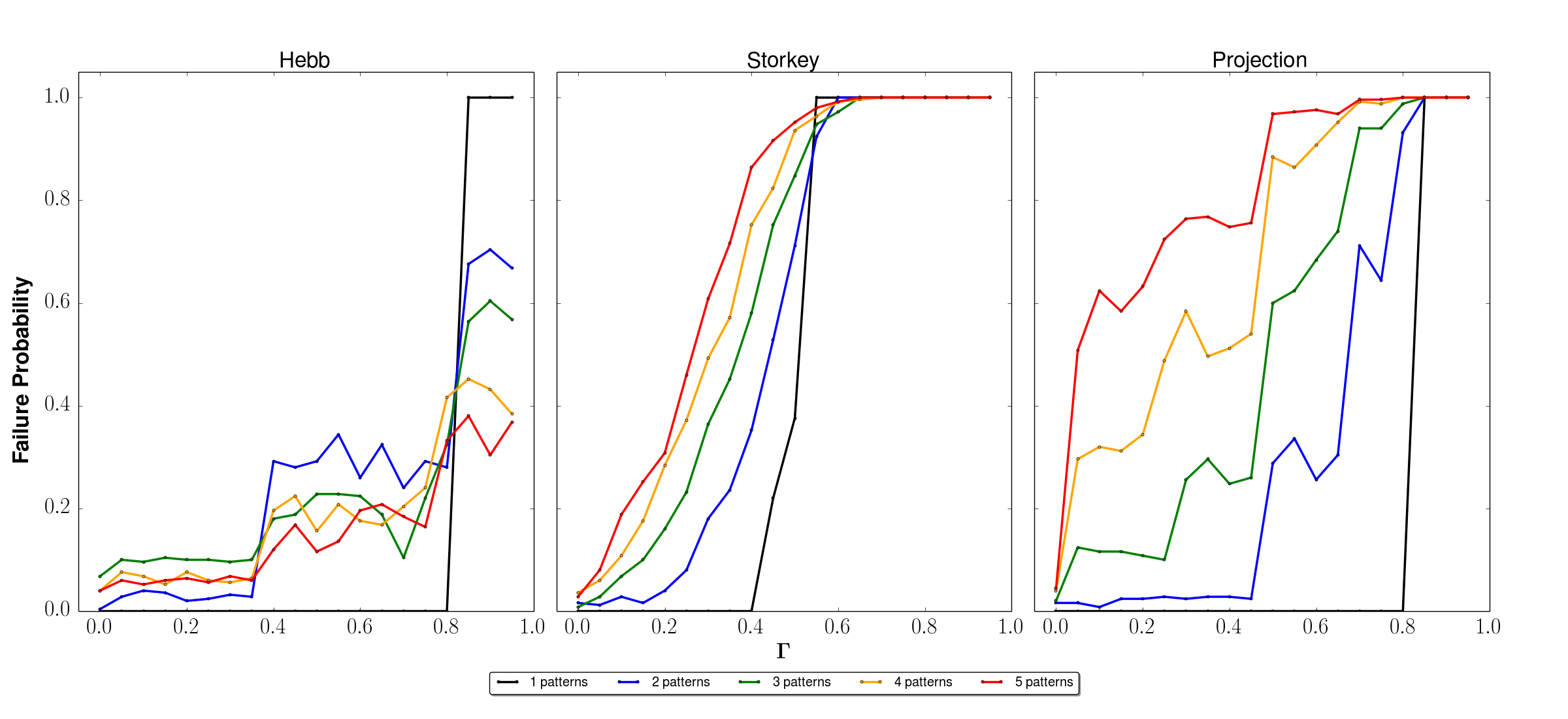}
\end{center}
\textbf{\refstepcounter{figure}
\label{fig:exp4_n5_hd1}Figure \arabic{figure}. }{Average probability to recall a memory not stored in the network with respect to applied bias, i.e., failure. The average is taken over an ensemble of network instances with $n=5$ neurons and $p=1,2,3,4$, or $5$ stored memories. Each panel plots the failure probability of a learning rule (Hebb, Storkey, projection) using an input state that is Hamming distance 1 from a stored memory. Failure increases with $\Gamma$ when the non-memory input state forms a fixed point in the network.}
\end{figure}
\begin{figure}
\begin{center}
\includegraphics[width=17cm]{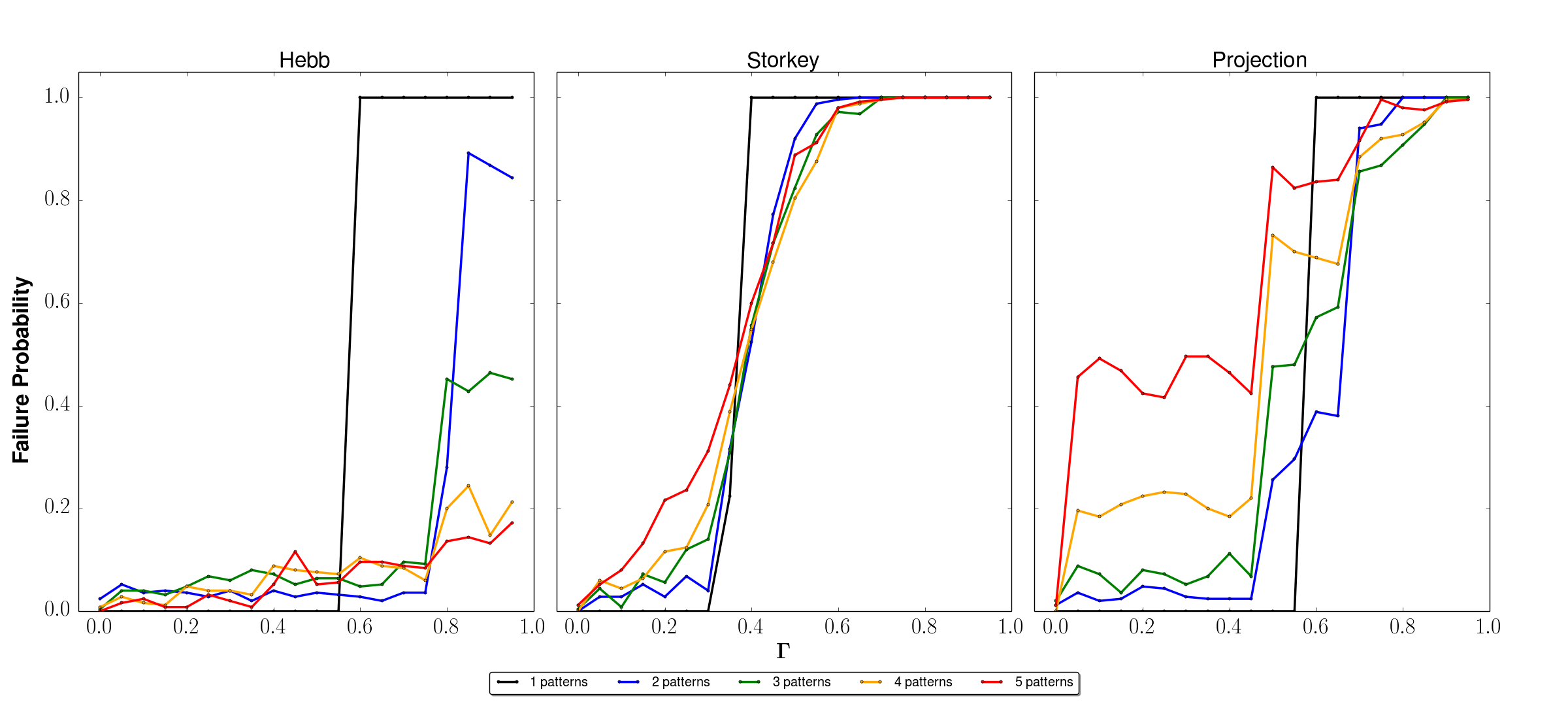}
\end{center}
\textbf{\refstepcounter{figure}
\label{fig:exp4_n5_hd2}Figure \arabic{figure}. }{Average probability to recall a memory not stored in the network with respect to applied bias, i.e. failure. The average is taken over an ensemble of network instances with $n=5$ neurons and $p=1,2,3,4$, or $5$ stored memories. Each panel plots the failure of a learning rule (Hebb, Storkey, projection) using an input state that is Hamming distance 2 from a stored memory. Failure increases with $\Gamma$ when the non-memory input state forms a fixed point in the network.}
\end{figure}

We have investigated further the influence on the network of over-biasing the input state. As noted previously, there are loose upper bounds on $\Gamma$ based on the energetic analysis of the learning rules \cite{Neigovzen2009}. We have tested these bounds by attempting to recall a memory that is not stored in the network. These tests attempt to recall a stored memory using an input state that is guaranteed to be either Hamming distance 1 or 2 away from any stored memory. We would expect the failure rate for this test to increase as either the number of stored patterns or the bias $\Gamma$ increases.  This is because the biased state should eventually reach an energy lower than any stored memory. Figure \ref{fig:exp4_n5_hd1} plots the average failure as the recall accuracy $\langle f_x \rangle $. In these plots, the input state is not among the stored memories. As expected, the failure rate increases as $\Gamma$ increases. For all the learning rules, there is a narrow range for $\Gamma$ above which the network returns the input state. These thresholds mark that the system is over biased. It is notable, however, that each learning rules exhibit a different behavior from over biasing. Whereas the Hebb rule terminates at lower failure probability as the memories are increased, both the Storkey and projection rules reach unit failure with sufficiently large $\Gamma$. This is again due to the inability for the Hebb rule to discriminate between interfering memories. A similar plot is shown in Fig.~\ref{fig:exp4_n5_hd2} for the case that the input is at least Hamming distance 2 from all the stored memories. The sensitivity to failure increases with the increase in Hamming distance as noted by the lower thresholds on $\Gamma$ for over biasing.  
\par
Finally, we have investigated the role of the annealing time $T$ on recall success. Because the state dynamics must be adiabatic relative to the minimum energy gap, the diversity of instances used for $\langle f_x \rangle$ are also likely to support a diversity of $\Delta_{min}$. This implies that there may be some maximum $T$ for the ensemble which ensures every instance is quasi-adiabatic. In Fig.~\ref{fig:fx_vs_p_multi}, we show a series of recall averages for different annealing times. For small values of $T$, the average success is low, especially as $p$ approaches $n$. This suggests that many instances do not meet the $x = 2/3$ threshold for success. As $T$ increases, the average success also increases but only up to a limit that depends on each learning rule. For the Storkey and projection rule, this limit is before $T = 500$, while for the Hebb rule the limit occurs before $T =50$. Annealing times larger than these limits do not lead to significant changes in the average recall success (assuming a linear annealing schedule). Thus the annealing time is not the limiting factor in recall success and the adiabatic condition has been met for these problem instances. Notably, the shortest annealing time is found for the Hebb rule but this rule also exhibits the most interference during memory recall. By contrast, the projection and Storkey rule require an order of magnitude increase in annealing time to ensure adiabaticity but these rules also exhibit greater accuracy during memory recall.
%%%%%%%%%%
%%%%%%%%%%
\section{Conclusions}
\label{sec:con}
We have presented a theoretical formulation of auto-associative memory recall in terms of adiabatic quantum optimization. We have used numerical simulations to quantify the recall success with respect to three different learning rules (Hebb, Storkey, and projection) and we have accumulated statistics on recall accuracy and failure across an ensemble of different network instances. We have found that the probability to populate the expected ground state using AQO is sensitive to learning rule, number of memories, and size of the network. Our simulation studies have been limited in size, but for these small networks there are notable differences in both the success and failure rates across learning rules. These differences represent the strategies of each learning rule to manage memory interference and the sensitivity of the AQO algorithm to those different strategies.
\par
The use of AQO for memory recall is closely related to its use for searching an unsorted database \cite{Farhi2000,Roland2002}. Both Farhi et al.~and Roland and Cerf have previously constructed the search problem using an oracle based on projection operators, which with an unbiased Hopfield network trained using the Hebb rule. Their previous work considered the task of recovering any valid memory from the network. We have used the activation threshold $\theta$ of the Hopfield network as the input key for a context-addressable memory. The activation threshold corresponds to the classical input to the oracle that identifies which memory is being sought. In this sense, the Hopfield network offers a robust implementation of Grover's search by permitting input to the task. However, this comes at the cost of a more complex oracle implementation. The three learning rules discussed here represent three different methods for oracle construction within the model of an Ising Hamiltonian. We have shown how choices in learning rule impact recall accuracy and we have observed that the projection rule seems to offer the most robust behavior. We have not attempted to optimize the annealing schedule associated with memory recall for each learning rule. It seems unlikely that the optimized annealing schedule recovered by Roland and Cerf for untagged search would extend to the current oracle implementations due to the influence of the variable activation threshold.
\par
Recent work to assess the scaling of the spectral gap that determines the minimum AQO annealing time has underscored that the relative height of energy barriers play a fundamental role in determining which Ising Hamiltonians are challenging \cite{Boixo2012, Karimi2012, Ronnow2014, Katzgraber2014}. Historically, learning rules that provide well separated but broad energy basins have been the goal of classical Hopfield networks, as these landscapes favor methods like gradient descent \cite{Hopfield1982, Personnaz1986, Storkey1999}. We have found that the AQO recall accuracy and minimal annealing time also demonstrate a significant dependence on the learning rule. In particular, energy basins prepared by the projection rule are known to be better separated than by either the Hebb or Storkey rules. Consequently, the projection rule provides the best performance with respect to AQO recall accuracy. However, better performance is not due to the avoidance of local minima but rather to the reduced interference between the stored memories and the biased input. Because the shape of the energy basins also influence the spectral gap of the time-dependent Hamiltonian, we anticipate that learning rules can provide a form of energetic control over AQO scaling.
%%%%%%%%%%
\section*{Acknowledgments}
H.~S.~thanks the Department of Energy Science Undergraduate Laboratory Internship (SULI) program and Prof.~Mark Edwards, Georgia Southern University, for use of computing resources. This manuscript has been authored by a contractor of the U.S. Government under Contract No.~DE-AC05-00OR22725. Accordingly, the U.S. Government retains a non-exclusive, royalty-free license to publish or reproduce the published form of this contribution, or allow others to do so, for U.S. Government purposes.

%%%%%%%%%%%
%% BIBLIOGRAPHY
%%%%%%%%%%%
\section*{References}
\bibliographystyle{ieeetr}
\bibliography{iop_manuscript.bibliography}

\section*{Figures}

\end{document}